# Negative Ion Sources: Magnetron and Penning

*D.C. Faircloth*

Rutherford Appleton Laboratory, Chilton, Oxfordshire, UK

**Abstract**

The history of the magnetron and Penning electrode geometry is briefly outlined. Plasma generation by electrical discharge-driven electron impact ionization is described and the basic physics of plasma and electrodes relevant to magnetron and Penning discharges are explained. Negative ions and their applications are introduced, along with their production mechanisms. Caesium and surface production of negative ions are detailed. Technical details of how to build magnetron and Penning surface plasma sources are given, along with examples of specific sources from around the world. Failure modes are listed and lifetimes compared.

## 1 Introduction

### 1.1 Overview

Magnetron and Penning ion sources have a long history going back to the early part of the 20th century. Their use for negative ion production really took off in 1970s Soviet Russia with the introduction of caesium. Caesiated magnetron and Penning negative ion sources are referred to as surface plasma sources. They are the brightest of all negative ion sources, with H⁻ beam current densities at extraction exceeding $1~\text{A}~\text{cm}^{-2}$.

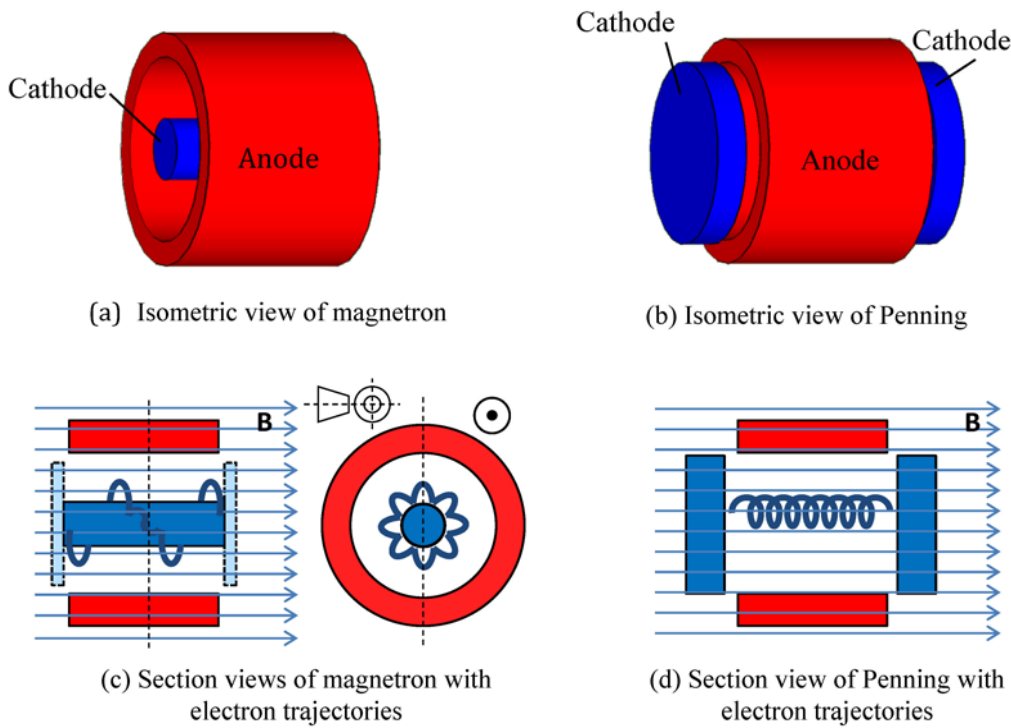

**Fig. 1:** Different views of the basic magnetron and Penning electrode geometries

## 1.2 Basic electrode topology

The fundamental magnetron and Penning topologies are shown in Fig. 1. Topologically, they both consist of a tube-shaped anode with a magnetic field running along the axis of the tube. The magnetron has a cathode in the centre of the tube as shown in Fig. 1(a), whereas the Penning has two cathodes at either end of the tube as shown in Fig. 1(b).

Although Fig. 1 shows the fundamental topologies, actual geometries used in practice are often quite different from those shown in Fig. 1. Anode tubes can be very short, so, instead of a tube, a window frame could be a better description of the anode shape. Electrodes can be square rather than round, they can be stretched to be rectangular, and often include a range of extra features such as lips, ridges, notches, grooves and dimples. Magnetron cathodes are often spool-shaped as it is shown in Fig. 1(c).

Whatever the geometry details, the basic topology of a magnetic field running perpendicularly through a hole in an anode is fundamental to both magnetron and Penning. It is the cathode topology and the resultant motion of the electrons (see section 2.4) that differentiates between them.

## 1.3 Early history

### 1.3.1 Magnetron

The magnetron electrode topology was first invented, not as an ion source, but as an electron tube by Albert Hull working at the General Electric Research Laboratory, Schenectady, New York in 1916 [1]. He was trying to use magnetism to find alternatives to the patented electrostatic control of valve amplifiers. The tubes he invented had various names, such as comet valves, boomerang valves and ballistic valves, because of the electron trajectories in the tube (see Fig. 1(c)) – however, the name magnetron valve eventually stuck [2].

To reach higher and higher powers, Hull started to add various gases to his tubes. By 1921 an experimental 1000 kW water-cooled magnetron with an anode voltage of 20 kV had been developed [2]. At the time Hull was working for the influential Irvin Langmuir. Langmuir discussed Hull's results with his fellow New England researchers and suggested that the gaseous discharge magnetron could make an excellent source of ions for the early particle accelerators being developed to study nuclear processes. The magnetron ion source was initially developed by Lamar and Luhr in the early 1930s at the Massachusetts Institute of Technology [3]. In 1934, Stanley Van Voorhis and his team in Princeton were the first to report using a magnetron as a proton source for an accelerator [4].

The possibility of using the magnetron electrode geometry to generate high-power microwaves was first practically demonstrated in 1940 by two British physicists, Henry Boot and John Randall working at Birmingham University, UK. By adding circular caves or cavities into the inner walls of the anode tube, electrons can be made to circulate resonantly in the cavities at microwave frequencies. With the addition of an output coupler positioned in one of the anode cavities, it is possible to produce high-power microwaves. These types of devices are known as cavity magnetrons and are not the subject of this paper. Cavity magnetrons are in common usage in microwave ovens.

### 1.3.2 Penning

The Penning geometry was first reported as a positive ion source by Louis Maxwell working at the Franklin Institute in Philadelphia in 1930 [5]. However, the Penning electrode geometry gets its name from Frans Penning, a researcher working at Philips Physics Laboratory in Eindhoven, The Netherlands. In 1937 he developed the Penning ionization gauge or Philips ionization gauge (PIG) [6]. By measuring the discharge current in the electrode geometry, he was able to measure accurately and reliably very low pressures in gases, a technique that is still widely used today.

*1.3.3   Variations*

The success of the magnetron and Penning electrode geometries as ion sources spawned a series of variations: the magnetron source was modified to give the Freeman source; the Penning source developed into the Calutron, Bernas and Nielson sources. Within a decade it was possible to produce multi-milliamp beams of positive ions from almost all the elements, no matter if they were solids, liquids or gases.

## 2   Plasma generation by electrical discharge

### 2.1   Introduction

An electrical discharge in the magnetron and Penning electrode configurations generates a plasma from which the required ions can be extracted. A low-pressure (< 1 mbar) gas is fed into the space between the electrodes. The type of gas used depends on the type of ions required. The electrical discharge is created by applying an electric potential between the anode and cathode. The resultant electric field accelerates free electrons, which can ionize the gas molecules. This process is called electron impact ionization.

### 2.2   Electrical discharges

*2.2.1   Townsend breakdown*

Electron impact ionization occurs when the applied electric field is high enough to accelerate the electrons (within their mean free path) to the ionization energy of the gas. At this point the current between the electrodes rapidly increases. The electrons ionize the neutral atoms and molecules, producing more electrons. These additional electrons are accelerated to ionize even more atoms, producing even more free electrons in an avalanche breakdown process known as Townsend breakdown. This runaway process means that the voltage needed to sustain the discharge drops significantly as the discharge enters the glow regime.

*2.2.2   Glow discharge*

A glow discharge is so called because it emits a significant amount of light. Most of the photons that make up this light are produced when atoms that have had their orbital electrons excited by electron bombardment relax back to their ground states. Photons are produced in any event that needs to release energy, for example when ions recombine with the free electrons or when vibrationally excited molecules relax.

A glow discharge is self-sustaining because positive ions that are accelerated to the cathode impact on it, producing more electrons in a process called secondary emission. This is why the voltage at which a discharge goes out is lower than the voltage at which it initiates.

The current in a glow discharge increases with very little increase in discharge voltage. The plasma distributes itself around the cathode surface as the current increases. Eventually the current reaches a point where the cathode surface is completely covered with plasma and the only way to increase the current further is to increase the current density at the cathode. This causes the plasma voltage near the cathode to rise.

*2.2.3   Arc discharge*

The increased current density leads to cathode heating, and eventually the cathode surface reaches a temperature where it starts to thermionically emit electrons (see section 2.3.9) and the discharge moves into the arc regime with a negative current versus voltage characteristic.

Magnetron and Penning ion sources operate with discharges in the glow arc regime. Technically the discharge is not an arc because the cathode is kept cool enough so it does not thermionically emit many electrons. However, the addition of caesium (see section 4) causes significant electron emission, which also yields a negative current versus voltage characteristic. The discharge current in magnetron and Penning ion sources is often called the 'arc current'.

## 2.3   Plasma properties

### 2.3.1   *Introduction*

The properties of the plasma produced in both the magnetron and Penning discharge share a lot of similarities. This section describes some basic plasma and electrode properties that will be referred to in later sections of this chapter.

### 2.3.2   *Density, n*

The most basic parameter is the density of each of the constituents in the plasma. It is usually written as $n$ with a subscript to represent the type of particle, and is expressed in number of particles per cubic metre. Some older papers give density in particles per cubic centimetre. Thus we have

$n_e$ = density of electrons
$n_i$ = density of ions
$n_n$ = density of neutrals

### 2.3.3   *Temperature, T*

The temperature of the plasma is a measurement of how fast each of the particles is going, otherwise known as the particle's kinetic energy. Particle temperatures are usually expressed in electronvolts. The Boltzmann constant $k_B$ makes 1 eV = $1.602 \times 10^{-19}/1.381 \times 10^{-23}$ = 11 600 K. Similarly we have

$T_e$ = temperature of electrons
$T_i$ = temperature of ions
$T_n$ = temperature of neutrals

The temperatures of the electrons, ions and neutrals can be different.

### 2.3.4   *Temperature distributions*

Obviously not all electrons in plasma will have the same temperature, and the same is true for the ions of the same species. The temperatures $T_e$, $T_i$ and $T_n$ are merely averages (of absolute temperatures). If the plasma is in thermal equilibrium, then the distribution will be Maxwellian and obey Maxwell–Boltzmann statistics.

Often the plasma is in a magnetic field. The ions and electrons will spiral around the magnetic field lines and slowly move along them. Hence, the particle velocities (temperatures) will not be the same in all directions. The particle temperatures in different directions are defined as follows: $T_{i\parallel}$ is the ion temperature parallel to the magnetic field, and $T_{i\perp}$ is the ion temperature perpendicular to the magnetic field.

### 2.3.5   *Quasi-neutrality*

Plasma is generally charge neutral, so all of the charge states of all the ions add up to the same number as the number of electrons:

$$\sum q_i\, n_i = n_e \tag{1}$$

### 2.3.6 Percentage ionization

The percentage ionization is a measure of how ionized the gas is, i.e. what proportion of the atoms have actually been ionized:

$$\text{percentage ionization} = 100 \times \frac{n_\text{i}}{n_\text{i}+n_\text{n}} \qquad (2)$$

When the percentage is above 10%, the plasma is said to be highly ionized and the interactions that take place within are dominated by plasma physics. Less than 1% ionization and interactions with neutrals must be considered.

### 2.3.7 Collisions

Collisions between charged particles in a plasma are fundamentally different from collisions in a neutral gas. The ions in a plasma interact by the Coulomb force: they can be attracted or repelled from a great distance. As an ion moves in a plasma, its direction is gradually changed as it passes the electric fields of neighbouring particles; whereas, in a neutral gas, the particles only interact when they get so close to each other that they literally bounce off each other's outer electron orbitals. In a neutral gas, the average distance the particles travel in a straight line before bouncing off another particle is referred to as the mean free path. In a plasma, the mean free path concept does not work because the ions and electrons are always interacting with each other by their electric fields. Instead, a concept called 'relaxation time' is invoked: this is the time it takes for an ion to change direction by 90°. The relaxation time $\tau_\theta$ can also be described as 'the average 90° deflection time'. In a plasma there are different relaxation times between each of the different particle species.

### 2.3.8 Work function, φ

In any solid metal, there are one or two electrons per atom that are free to move from atom to atom. This is sometimes collectively referred to as a 'sea of electrons'. Their velocities follow a statistical distribution, rather than being uniform. Occasionally an electron will have enough velocity to exit the metal without being pulled back in. The minimum amount of energy needed for an electron to leave a surface is called the work function. Specifically, the work function, $\varphi$, shown in Fig. 2, is the energy needed to move an electron from the Fermi level $E_\text{f}$ into vacuum $E_0$. 'Fermi level' is the term used to describe the top of the collection of electron energy levels in an atom at absolute zero temperature. The work function is characteristic of the material surface and for most metals is of the order of several electronvolts.

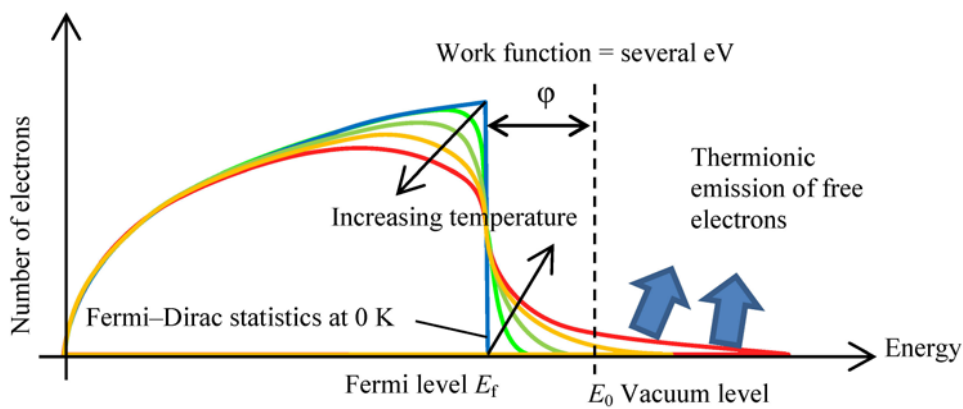

**Fig. 2:** Electron energy distribution in metals at different temperatures and the work function

### 2.3.9 Thermionic emission

Thermionic emission is the heat-induced flow of charge carriers from a surface or over a potential energy barrier. This occurs because the thermal energy given to the carrier overcomes the work function of the metal. Heating a material causes the energy distribution of the electrons to spread out, as shown in Fig. 2. Thermionic currents can be increased by decreasing the work function. This often-desired goal can be achieved by applying various oxide coatings to the surface.

In 1901 Owen Richardson found that the current from a heated wire varied exponentially with temperature. He later proposed this equation:

$$J = A_\text{G} T^2 e^{\frac{-\varphi}{kT}} \quad (3)$$

where $J$ is the electron current density on the surface of the cathode, $\varphi$ is the cathode work function and $T$ is the temperature of the cathode. Here $A_\text{G}$ is given by

$$A_\text{G} = \lambda_\text{R} A_0 \quad (4)$$

where $\lambda_\text{R}$ is a material-specific correction factor that is typically of order 0.5 and $A_0$ is a universal constant given by

$$A_0 = \frac{4\pi m k^2 e}{h^3} = 1.20173 \times 10^6 \text{ A m}^{-2} \text{ K}^{-2} \quad (5)$$

where $m$ is the electron mass, $e$ is the elementary charge and $h$ is Planck's constant.

### 2.3.10 Debye length

Named after the Dutch scientist Peter Debye, the Debye length, $\lambda_\text{D}$, is the distance over which the free electrons redistribute themselves to screen out electric fields in plasma. This screening process occurs because the light mobile electrons are repelled from each other while being pulled by neighbouring heavy, low-mobility positive ions. Thus the electrons will always distribute themselves between the ions. Their electric fields counteract the fields of the ions, creating a screening effect. The Debye length not only limits the influential range that particles' electric fields have on each other but also limits how far electric fields produced by voltages applied to electrodes can penetrate into the plasma. The Debye length effect is what makes the plasma quasi-neutral over long distances.

The higher the electron density, the more effective the screening, and thus the shorter this screening (Debye) length will be. The Debye length is given by:

$$\lambda_\text{D} = \sqrt{\frac{\epsilon_0 k T_\text{e}}{n_\text{e} q_\text{e}^2}} \quad (6)$$

where

$\lambda_\text{D}$ is the Debye length (of order 0.1–1 mm for ion source plasmas),

$\epsilon_0$ is the permittivity of free space,

$k$ is the Boltzmann constant,

$q_\text{e}$ is the elemental charge,

$T_\text{e}$ is the temperature of the electrons,

$n_\text{e}$ is the density of electrons.

### 2.3.11 Plasma sheath

The screening effect of the plasma creates a phenomenon called the plasma sheath around the cathode electrode(s). The plasma sheath is also called the Debye sheath. The sheath has a greater density of positive ions, and hence an overall excess positive charge. It balances an opposite negative charge on the cathode with which it is in contact. The plasma sheath is several Debye lengths thick.

The quasi-uniform plasma potential is closest to the anode voltage, so the largest potential drop in a plasma is across the plasma sheath near the cathode. Positive ions that drift into the cathode sheath region are accelerated towards the cathode by this potential drop. Similarly, negative ions and electrons produced on the cathode surface are accelerated away.

## 2.4 Magnetic field and electron trajectories

Although the magnetron and Penning discharge geometries share a lot of the basic plasma properties listed in the previous section, they have fundamentally different electron trajectories. This is caused by the different way their electric and magnetic fields are oriented.

Figure 3(c) shows the direction of the dipole magnetic field in the magnetron geometry. Without the magnetic field, the electrons would simply flow axially from the cathode to the anode, as shown in Fig. 3(a). The magnetic field causes the electrons emitted from the cathode to rotate and end up back near the cathode. They are repelled from the cathode only to be rotated back towards the cathode again. This process repeats, causing the electrons to propagate around the cathode in a looping circular motion. As well as propagating around the cathode, the electrons also drift along its length in a helix. The magnetron cathode usually has end discs that make the cathode resemble a spool. The purpose of the end discs is to repel the electrons when they reach the end of the cathode. In this way the electrons spiral up and down the spool-shaped cathode.

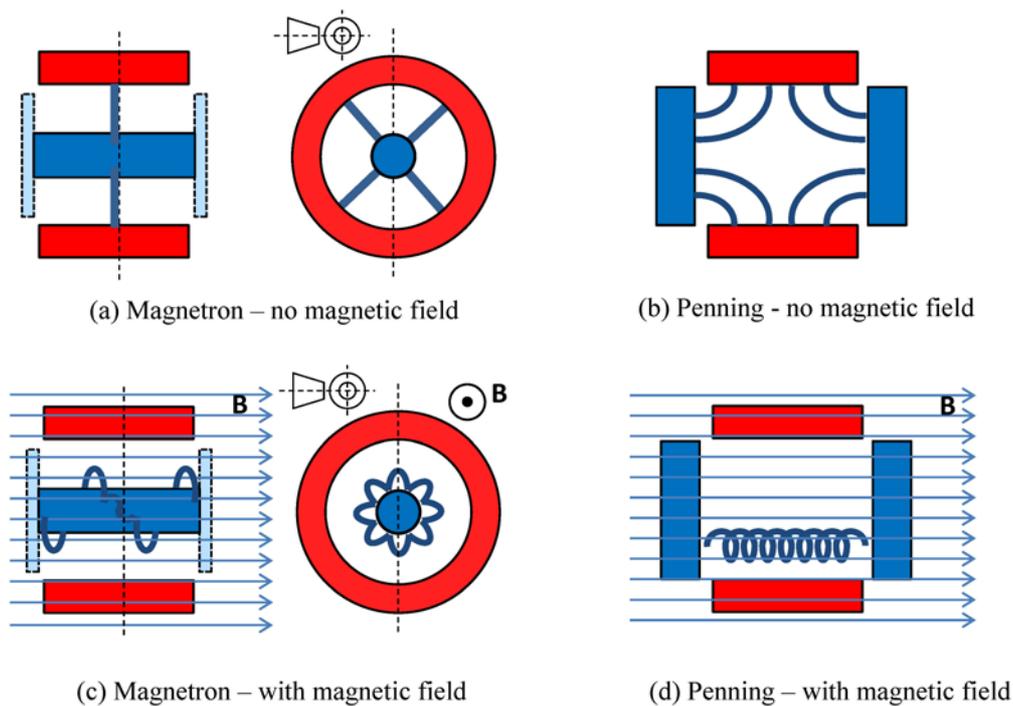

(a) Magnetron – no magnetic field    (b) Penning - no magnetic field

(c) Magnetron – with magnetic field    (d) Penning – with magnetic field

**Fig. 3**: Electron trajectories in magnetron and Penning electrode geometries with and without magnetic fields

Figure 3(d) shows the direction of the dipole magnetic field in the Penning geometry. Without the magnetic field, the electrons would follow the quadrupole-shaped electric field lines, as shown in Fig. 3(b). With the magnetic field, the electrons are emitted from a cathode, and then they drift along the magnetic field lines by spiralling around them. They eventually get to the opposite cathode, where they are reflected back.

In reality, the electron trajectories are much more complex than the idealized ones that are shown in Fig. 3. They have a range of energies (section 2.3.4) and they interact with the electric fields of the ions and other electrons (section 2.3.7), leading to effects such as screening (section 2.3.10) and the cathode sheath (section 2.3.11). However, the overall effect is that the magnetic field confines the electrons: to circulate around the cathode in the case of the magnetron, and to bounce between the two cathodes in the Penning.

# 3   Negative ions

## 3.1   The negative ion

Negative ions are neutral atoms with an additional electron attached. The binding energy of the additional electron to the atom is termed the electron affinity. Some elements (such as beryllium, nitrogen and the noble elements) have a negative electron affinity, which means that they cannot form stable negative ions. Both magnetron and Penning ion sources have been used to produce $H^-$ and $D^-$ ions as well as other heavy negative ions, such as $O^-$, $B^-$, $C^-$, etc.

$H^-$ ions are the most commonly produced negative ion. Hydrogen has an electron affinity of 0.7542 eV. Considering that the binding energy of an electron to a proton is 13.6 eV, the extra electron on an $H^-$ ion is very loosely held on.

It is also important to note that $H^-$ has a much larger ionization cross-section than $H^0$: 30 times larger for electron collisions and 100 times for $H^+$ collisions. $H^-$ ions are very fragile.

## 3.2   Uses

Negative ion sources were first developed to allow electrostatic accelerators to effectively double their output beam energy. In a tandem generator, $H^-$ ions are first accelerated from ground to terminal volts; they are then stripped of their two electrons when they pass through a thin foil. The resulting protons are then accelerated from terminal volts back to ground, at which point they have an energy of twice the terminal volts.

Cyclotrons use negative ions and stripping foils to extract the beam from the cyclotron. The stripping foil is positioned near the perimeter of the cyclotron poles. As the negative ion beam is accelerated, it circulates on larger and larger radii until it passes through the stripping foil, which converts the beam from being negative to positive. The Lorenz force on the beam is reversed and, instead of the force pointing into the centre of the cyclotron, it points outwards and the beam is cleanly extracted.

In high-power proton accelerators, $H^-$ ions are used to allow charge accumulation via multiturn injection. An $H^-$ beam from a linear accelerator is fed through a stripping foil into a circular ring (a storage, accumulator or synchrotron ring), leaving protons circulating in the ring. The $H^-$ beam from the linear accelerator continues to enter the ring while the circulating beam repeatedly passes through the stripping foil unaffected. The incoming beam curves in one way through a dipole as an $H^-$ beam, then curves out of the dipole in the opposite direction as a proton beam on top of the circulating beam. This allows accelerator designers to beat Liouville's theorem and build up charges in phase space. Without this negative ion stripping trick, only one turn on the same orbit could be accumulated in the ring.

## 3.3   Early negative ion sources

### 3.3.1   Charge exchange

The first $H^-$ ion sources were charge exchange devices. There are two ways of doing this: with foils or gases. With foils, a proton beam, with an energy of about 10 keV, is passed through a negatively biased foil and by electron capture an $H^-$ beam is produced. For gases, the proton beam is passed through a region filled with a gas. The $H^-$ beam is produced by sequential electron capture; protons are converted first to neutral H⁰, then to H⁻. The gas acts as an electron donor. Only about 2% of the protons are converted into H⁻ ions. Until the 1960s this was the main technique used to make H⁻ beams. Beams of up to 200 µA were produced using this method. In 1967 Bailey Donnally [7] discovered that the yield of He⁻ ions can be increased by using caesium vapour as an electron donor. This led to the development of a series of negative ion sources using alkali vapour.

### *3.3.2 Extraction from the edge of the plasma*

For several decades, numerous researchers [8, 9] had been experimenting with sources originally designed to produce positive ions, but by reversing the polarity of the extraction they were able to extract negative ions. Nevertheless, the co-extracted electron current was always at least an order of magnitude higher than the negative ion current.

In the early 1960s George Lawrence and his team at Los Alamos [10] were using a duoplasmatron to produce H⁻ ions when they first noticed that substantially higher beam currents and lower electron currents could be extracted when the extraction hole was moved away from the centre of the discharge (Fig. 4). They concluded that the extracted H⁻ ions must be produced near the edge of the plasma. (This was also discovered independently by a team at the UK Atomic Weapons Establishment [11].)

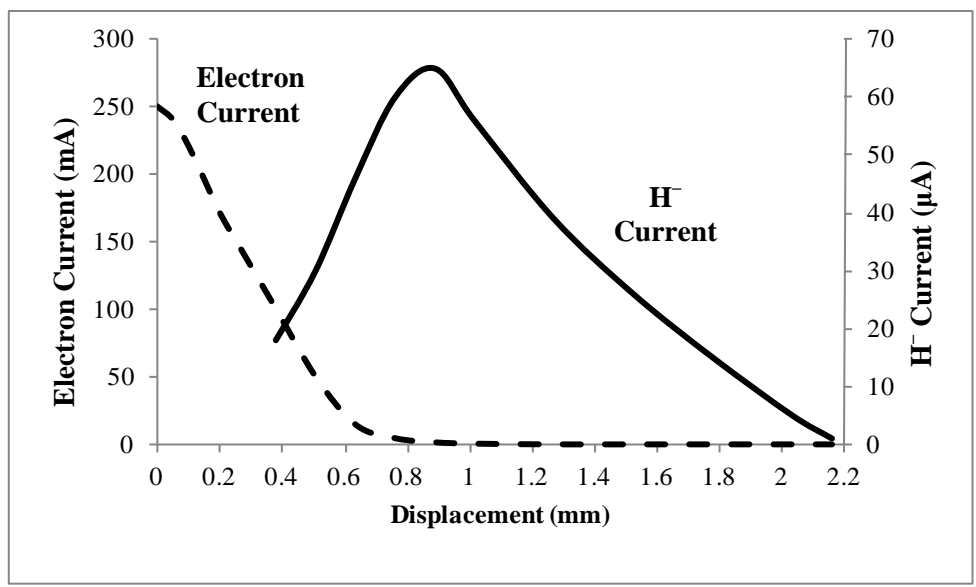

**Fig. 4**: H⁻ and electron currents as a function of extraction offset in a duoplasmatron measured at Los Alamos

During the 1960s various sources originally designed to produce positive ions were adapted and modified to produce H⁻ ions, and beam currents up to a few milliamps were produced.

### **3.4 Production mechanisms**

### *3.4.1 Overview*

The physical processes involved with the production of negative ions are still not fully understood, but they can be generally described as: charge exchange, surface and volume production processes. In different types of source, one production process may dominate; however, all three processes might contribute to the overall extracted negative ion current.

*3.4.2 Charge exchange*

Already described in section 3.3.1, charge exchange was the first method used to produce $H^-$ ions. Resonant charge exchange processes also occur in the Penning surface plasma source. $H^-$ ions produced on the cathode surface are accelerated by the cathode sheath (see section 2.3.11) to the plasma potential ($\approx 60$ V). If these fast $H^-$ were to be directly extracted they would create a beam with a large energy spread. The fast $H^-$ undergo resonant charge exchange with slow (thermal) neutral hydrogen atoms ($H^0$) in the plasma, leaving slow $H^-$ ions ready for extraction:

$$H^- (\approx 60 \text{ eV}) + H^0 (< 1 \text{ eV}) \rightarrow H^- (< 1 \text{ eV}) + H^0 (\approx 60 \text{ eV}) \qquad (7)$$

*3.4.3 Surface production*

In magnetron and Penning negative ion sources, the primary means of $H^-$ ion production is via caesium-enhanced surface production. This was discovered by Soviet scientists in the 1970s. They called ion sources using this production mechanism 'surface plasma sources' (SPS). Caesium-enhanced surface production of negative ions is discussed in more detail in section 4.

*3.4.4 Volume production*

Also in the 1970s Marthe Bacal working at Ecole Polytechnique discovered a completely new production mechanism that relied on $H^-$ production in the plasma volume itself. Initially people were sceptical because $H^-$ ions are so fragile (as discussed in section 3.1). The plasma in the discharge was thought to be too energetic for any $H^-$ ions produced in the volume to survive long enough to make it to the extraction region. The breakthrough that made volume production possible was separation by a magnetic filter field of the plasma production region from the extraction region. In the 1980s, Leung, Ehlers and Bacal used a filament-driven multicusp ion source with a magnetic dipole filter field positioned near the extraction region. The filter field blocked high-energy electrons from entering the extraction region, whereas ions and cold electrons could diffuse across the filter field. This effectively separated the discharge into two regions: a high-temperature driver plasma on the filament side of the filter field, and a low-temperature $H^-$ production plasma on the extraction region side. Magnetically filtered multicusp ion sources are sometimes called 'tandem' sources because of these two regions of different plasma temperatures (not to be confused with tandem accelerators).

The volume production process relies on the dissociative attachment of low-energy electrons to rovibrationally excited $H_2$ molecules:

$$H_2^* + e (\leq 1 \text{ eV}) \rightarrow H^- + H^0 \qquad (8)$$

If the $H_2$ molecule is vibrationally cold, the dissociative attachment cross-section is extremely low ($10^{-21}$ cm$^2$). When the $H_2$ molecule is rovibrationally excited, however, the cross-section increases by five orders of magnitude. Thus, low-energy electrons can be very effective in generating $H^-$ ions by dissociative attachment to highly vibrationally excited molecules. The rovibrationally excited molecules are produced not only in the plasma but also on the walls of the chamber and electrode surfaces.

It is possible that some of the extracted $H^-$ ions in the magnetron and Penning sources are created by volume production near the extraction aperture, with the magnetic dipole field acting as the fast electron filter.

**3.5 Negative ion extraction**

One of the challenges with negative ion source design is how to deal with the co-extracted electrons. Extracting electrons with the negative ions is obviously unavoidable because they both have the same charge. The ion source engineer must first try to minimize the amount of co-extracted electrons, and then find a way to separate and dump the unwanted electrons from the negative ion beam. The electron current can be 10 times greater than the negative ion current itself.

## 4 Caesium and surface production

### 4.1 History

#### 4.1.1 Caesium and surfaces

In the early 1960s Victor Krohn, Jr and his team [12] at Space Technology Laboratories, Inc., California were experimenting with surface sputter ion sources. Surface sputter ion sources are mainly used to produce beams of heavier ions (such as metals) for coating and etching applications. Krohn noticed that, when $Cs^+$ ions were used to sputter a metal target, the yield of sputtered negative ions increased by an order of magnitude.

#### 4.1.2 Soviet breakthrough

In the early 1970s Gennadii Dimov, Yuri Belchenko and Vadim Dudnikov at the Budker Institute of Nuclear Physics started experimenting with caesium in ion sources. Using a magnetron ion source (which the Soviets called a planotron), Dudnikov added Cs vapour to the discharge for the first time. A dramatic increase in $H^-$ current was observed, along with a decrease in co-extracted electrons. The Dimov team went on to extract a colossal 880 mA pulsed $H^-$ beam from an experimental magnetron ion source [13].

This success led them to develop a Penning-type ion source that could produce 150 mA of pulsed $H^-$ beam current with only 250 mA of extracted electrons [14]. The $H^-$ currents produced were orders of magnitude higher than anything seen previously. When these revolutionary results were published, interest in caesiated ion sources took off. Researchers all over the world started using caesium in their ion sources, and a large number of new ion source designs were developed.

The Soviets went on to develop a semi-planotron (essentially half a magnetron geometry where the electrons only circulate round half of a cathode spool), which they claimed was capable of delivering absolutely enormous 11 A $H^-$ beam currents.

#### 4.1.3 American developments

The most intense development occurred in the USA. Krsto Prelec and his team at Brookhaven National Laboratory (BNL) improved the geometry with the addition of a cathode indent (see section 5.3). They also developed a 2 A $H^-$ beam current magnetron for neutral beam injectors for fusion projects. Chuck Schmidt *et al.* took the BNL design to Fermilab and developed a magnetron for accelerator applications that still runs today.

At Los Alamos National Laboratory, Paul Allison *et al.* developed the Penning source design, which, combined with some aspects of the Fermilab design, was then used on the ISIS spallation neutron source in the UK.

A very different type of $H^-$ source that also relies on surface production of $H^-$ ions is the surface converter source. Developed in the 1980s by Ehlers and Leung at the Lawrence Berkeley Laboratory, it also relies on a caesiated surface [15]. A caesiated surface sits in the middle of a large filament-driven multicusp confined plasma. The surface is curved, with a radius centred on the extraction region. $H^-$ ions produced on this surface are geometrically 'focused' towards the extraction hole because they are repelled by a negative potential applied to the converter surface. This type of source is said to be 'self-extracting'.

## 4.2 Surface production of negative ions

### 4.2.1 Surface physics processes

Surfaces are very interesting places. Many different types of particles arrive at a surface: protons, ionized hydrogen molecules, ionized caesium, neutral atoms or molecules. They could be travelling at thermal velocities of a fraction of an electronvolt, or they could be travelling at up to 50–80 eV after being accelerated by the plasma sheath potential (section 2.3.11).

When a particle interacts with a surface, many complex and competing processes can occur:

- reflection,
- adsorption,
- sputtering,
- desorption,
- recombination,
- dissociation,
- ionization,
- neutralization,
- secondary electron emission,
- photoemission,
- excitation.

Particles ejected from the surface could be of a different charge state from the incoming particle, be excited, be in a molecule, or be some combination of all three. The surface may also be altered. Complex interactions can take place at the surfaces.

### 4.2.2 Low-work-function surface production

To make H⁻ ions, the surface must provide electrons. Surfaces tend to hold on to electrons – they have an electron affinity, or work function, $\varphi$. So to make H⁻ ions, a low-work-function surface is essential.

The work function of a surface obviously depends on what it is made of. If atoms of a different element are adsorbed on that surface (such as caesium), then the work function can be altered. The 'thickness' of the adsorbed layer will also have an effect on the surface's work function. The thickness of the adsorbed layer is usually defined in terms of the number of 'monolayers' of the adsorbed atoms:

$$\text{thickness (number of monolayers)} = \frac{\text{number of adsorbate atoms per unit area}}{\text{number of adsorbate atoms for a monolayer per unit area}} \qquad (9)$$

When talking about negative ion production, the surface is usually the cathode and is typically made of a high-melting-point metal such as tungsten, $\varphi = 4.55$ eV, or molybdenum, $\varphi = 4.6$ eV. Caesium has the lowest work function of all elements: $\varphi = 2.14$ eV. The work function of a caesium-coated molybdenum surface is actually lower than that of bulk caesium. As caesium covers the molybdenum surface, the work function decreases to 1.5 eV at 0.6 of a monolayer and then rises to about 2 eV for one monolayer or greater of caesium, as shown in Fig. 5. This minimum at 0.6 monolayers is caused by atomic interactions increasing the Fermi level at the surface, thus decreasing the amount of energy required to liberate the electrons.

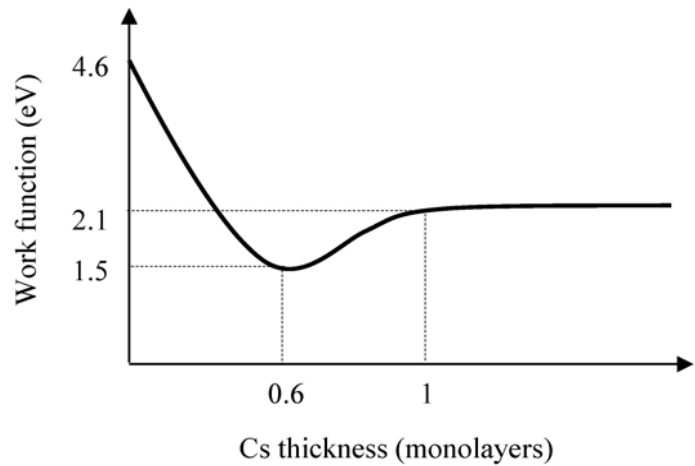

**Fig. 5:** Surface work function versus caesium thickness on a molybdenum surface

Although this sounds like an impossibly precise surface coverage to maintain, it is relatively simple to engineer. The thickness of the caesium coating on electrode surfaces will rarely ever come close to one full monolayer because thermal desorption and plasma sputtering remove excess caesium. This is because the Cs–Cs bond is weaker than the Cs–Mo, Cs–Ta or Cs–W bonds, so Cs atoms adsorbed on Cs atoms are rapidly sputtered away by the plasma or thermally emitted from hot surfaces. (It is possible to build up multiple layers but only on cold surfaces that are shielded from the plasma.)

*4.2.3   Maintaining caesium coverage*

To minimize the work function and maximize the H⁻ production, an optimum layer of caesium must be maintained on the surface. The surface is a dynamic place – caesium atoms are constantly being desorbed by plasma bombardment. To maintain optimum caesium coverage in a surface plasma source, a constant flux of caesium is required. This is often provided by an oven containing pure elemental caesium, but it can also be provided using caesium chromate cartridges that release Cs when heated.

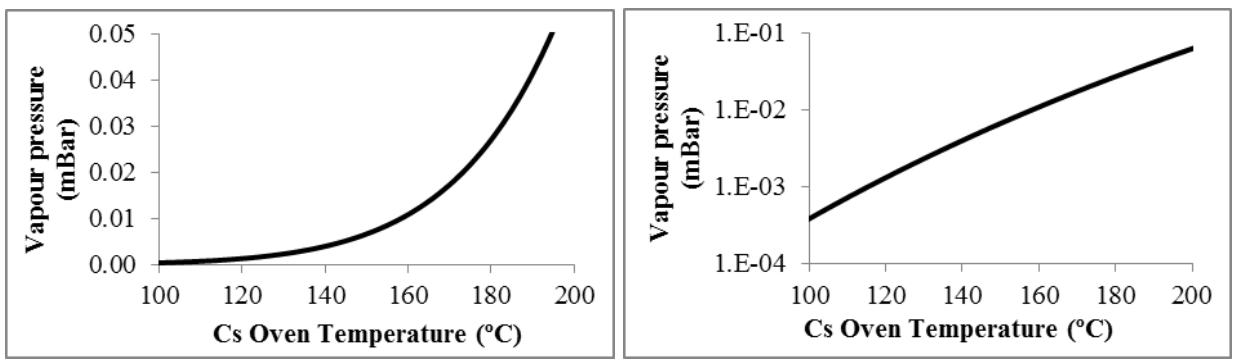

**Fig. 6:** Caesium vapour pressure versus caesium oven temperature

The flux of caesium into the plasma can be precisely controlled by setting the temperature of the caesium oven. Fig. 6 shows how the vapour pressure of caesium varies with temperature [16]. Ion sources that use elemental caesium in an oven usually operate somewhere between 70°C and 190°C. This covers a very large range of vapour pressures suitable for most applications.

### 4.3 H⁻ destruction

There are many processes that can destroy H⁻ ions. The most common are:

$$H^- + H^+ \rightarrow H^0 + H^0 \quad \text{mutual neutralization}$$
$$H^- + e \rightarrow H^0 + 2e \quad \text{electron detachment}$$
$$H^- + H^0 \rightarrow H_2^* + e \quad \text{associative detachment}$$

As already mentioned, the H⁻ ion cross-section is 30 times larger than the neutral H⁰ atom for collisions with electrons and 100 times larger for collisions with H⁺ ions. As well as being fragile and easily destroyed, it is much more likely to be hit.

The aim of the source designer is to minimize the H⁻ destruction processes by controlling the geometry, temperature, pressure and fields in the source.

### 4.4 The additional benefits of caesium

In addition to aiding H⁻ surface production, caesium also helps to stabilize the plasma by readily ionizing to produce additional electrons for the discharge. This reduces the amount of noise in the discharge and extracted beam current.

## 5 Magnetron surface plasma source

### 5.1 Introduction

Thus far this chapter has discussed features common to all magnetron and Penning electrode geometries, along with negative ions and surface production on caesiated surfaces. This section gives more details specifically about magnetron surface plasma negative ion sources. Examples from different operational particle accelerators around the world are given.

### 5.2 Construction

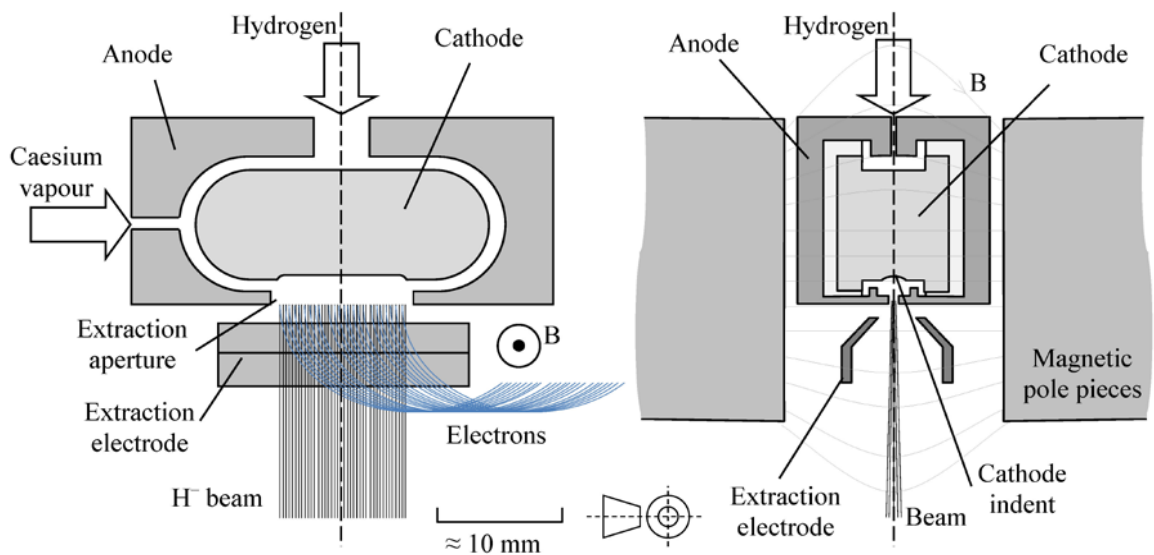

**Fig. 7:** Sectional schematic of a magnetron with slit extraction

The magnetron surface plasma source has a racetrack-shaped discharge bounded on the inside by the spool-shaped cathode and on the outside by the anode, as shown in Fig. 7. The anode and cathode are only about 1 mm apart, so the discharge is quite thin in the shape of a ribbon wrapped around the cathode. The anode and cathode are separated by a ceramic spacer. A magnetic field of between 0.1 T and 0.2 T is applied perpendicular to the plane of the racetrack, which causes the electrons to circulate around the racetrack. On one of the long sides of the racetrack discharge the anode has a hole through which the beam is extracted. Hydrogen is pulsed with a piezoelectric valve into the discharge on the opposite side to the extraction hole. Caesium vapour is introduced via an inlet on one side from a heated oven containing elemental caesium.

### 5.3 H⁻ production

There is an indent in the cathode opposite the extraction aperture. The indent increases the extracted H⁻ beam current by geometrically focusing the H⁻ ions produced on the cathode surface. The shape of the indent depends on the shape of the extraction geometry. A spherical dimple indent is used for a circular aperture and extraction electrode. A cylindrical groove indent is used for a slit aperture and extraction electrode, as shown in Fig. 6.

H⁻ ions are produced on the cathode surface and are accelerated away by the cathode sheath potential. Some of the cathode sheath-accelerated H⁻ ions undergo resonant charge exchange with slow thermal $H^0$ on the way to extraction, resulting in a beam energy distribution with two peaks.

### 5.4 Extraction

The amount of beam current extracted increases with extraction voltage. Voltages up to 35 kV have been used for ion sources on operational machines that produce 100 mA [17]. It is possible that the co-extracted electron current will be over 10 times the H⁻ beam current depending on how the magnetron is operated. There must be provision for dumping electron beam currents of this magnitude.

The ion source designer must consider the co-extracted electron current when specifying extraction power supplies. Caesium will also cause the extraction system to flash over in normal operation and during start-up. The high-voltage extraction power supply must be tough enough to withstand these breakdowns.

### 5.5 Temperature control

Technically speaking, the type of magnetron under discussion is a cold-cathode surface plasma magnetron. The term 'cold-cathode' is used because the cathode is not in the form of a heated filament. However, the name can be misleading because the cathode must still operate at elevated temperatures to maintain the Cs balance allowing stable and reliable operation. The electrodes generally run at a few hundred degrees Celsius; they are heated by the power of the discharge, and sometimes additional heating elements are used if the source operated at very low duty cycles. The temperature can be controlled by some form of cooling system if the source runs at higher duty cycles.

### 5.6 Caesium trapping

Caesium is exceptionally good at producing lots of electrons, so it can also promote high-voltage breakdowns if it gets onto electrode surfaces. The first accelerator applications of the negative magnetron used extra high voltages (EHV) generated by a Cockcroft–Walton voltage multiplier as the primary post-extraction acceleration technique. It was therefore essential to prevent caesium entering the high-voltage acceleration column by using some form of trap. The most common trap design is a 90° dipole inside a cold box, as shown in Fig. 8. The beam is bent through 90°, which also serves to separate out the co-extracted electrons; the beam exits through a hole in the cold box after the 90° bend. The cold box is held between −40°C and 0°C using refrigerant; this causes caesium vapour escaping from the source to condense on the walls of the cold box. The dipole field is also used in the ion source itself to confine the electrons. The refrigerated cold box is usually biased at the extraction electrode voltage so it can also be used to cool the extraction electrode by a direct thermal contact.

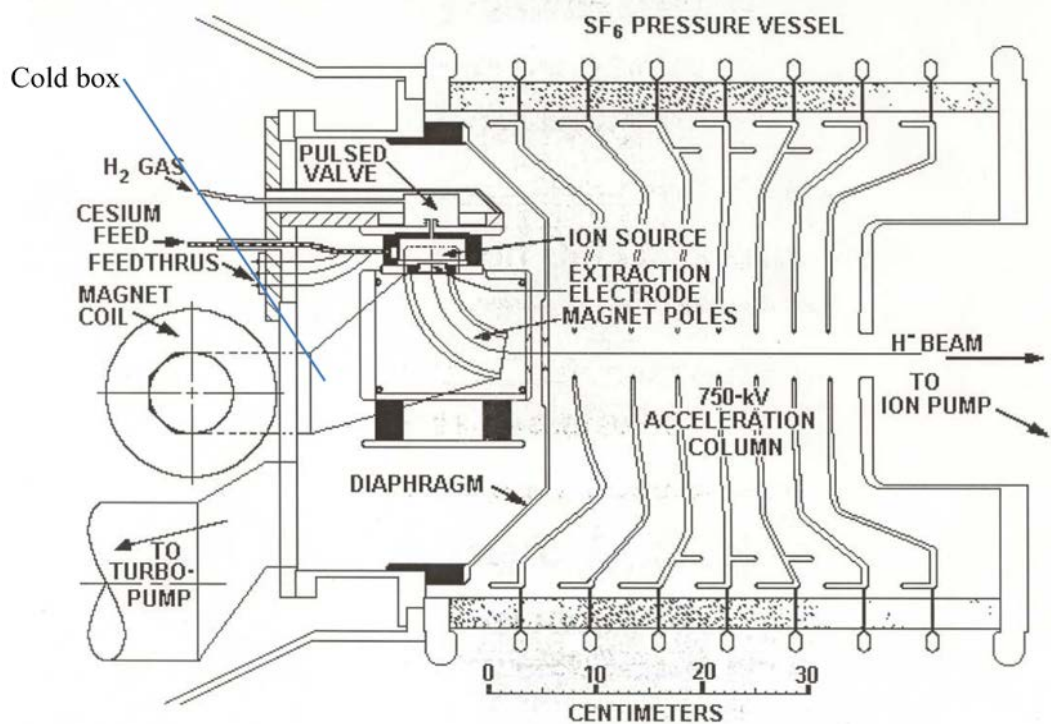

**Fig. 8:** The FNAL magnetron, cold box and 750 kV acceleration column

Radio frequency quadrupole (RFQ) accelerators have now largely replaced EHV. It is now common to post-extraction accelerate the beam up to somewhere between 30 and 70 keV, then pass the beam though a magnetic low-energy beam transport (LEBT) system into an RFQ. The LEBT is usually long enough (1–2 m) to prevent any caesium from entering the RFQ without a cold box trap. The caesium is deposited on the beam pipe walls in the LEBT, but this is less of a problem because only magnetic fields are present.

### 5.7 Duty cycle and noise limitations

The magnetron can deliver high $H^-$ currents of over 100 mA and can have very long lifetimes of over six months, but only at very low duty factors less than 0.5%. This because it is not possible to maintain optimum caesium coverage on the cathode surface opposite the extraction aperture for higher duty factors.

The beam from a magnetron is also noisier than that of a Penning ion source.

### 5.8 Examples of negative ion source magnetrons around the world

#### 5.8.1 *Budker Institute of Nuclear Physics (BINP), Novosibirsk, Russia*

The magnetron (also called a planotron) was the first ion source where the $H^-$ current was significantly increased by adding caesium vapour. This work was done by Gennadii Dimov, Yuri Belchenko and Vadim Dudnikov at the Budker Institute of Nuclear Physics in the early 1970s [13]. Many different magnetron designs were developed at BINP.

### 5.8.2 Fermi National Accelerator Laboratory (FNAL), USA

In the late 1970s Chuck Schmidt developed the FNAL version of the magnetron ion source shown in Fig. 9. The ion source is mounted face down on a flange inserted from the rear, as shown in Fig. 8. The negative ion beam is then extracted and bent through a 90° bend in a cold box caesium trap.

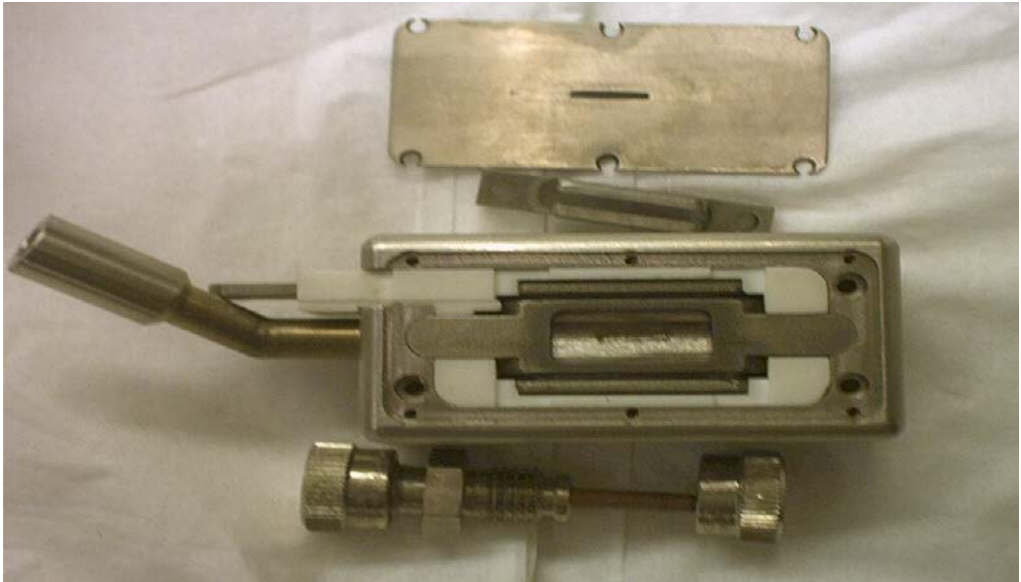

**Fig. 9:** The FNAL surface plasma magnetron source

### 5.8.3 Deutsches Elektronen-Synchrotron (DESY), Hamburg, Germany

The FNAL magnetron design was adopted by DESY and further developed by Jens Peters [18]. The DESY magnetron ran on the HERA accelerator. The design is slightly different from the FNAL design in that the ion source is top loaded into the 90° magnet and caesium trap cold box, as shown in Fig. 10.

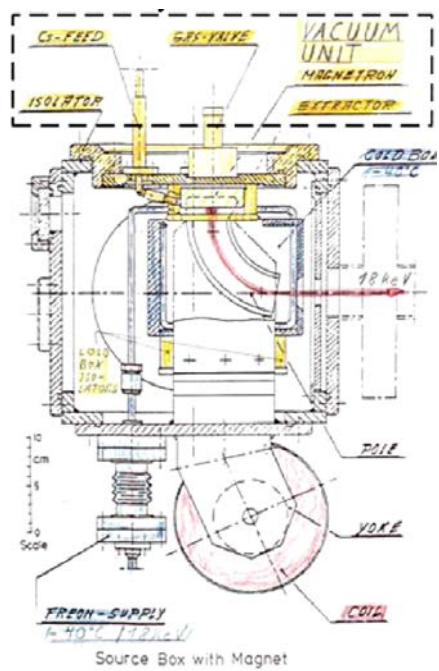

**Fig. 10:** The HERA top-loading magnetron and caesium trap

The different components of the HERA magnetron are shown in Fig. 11.

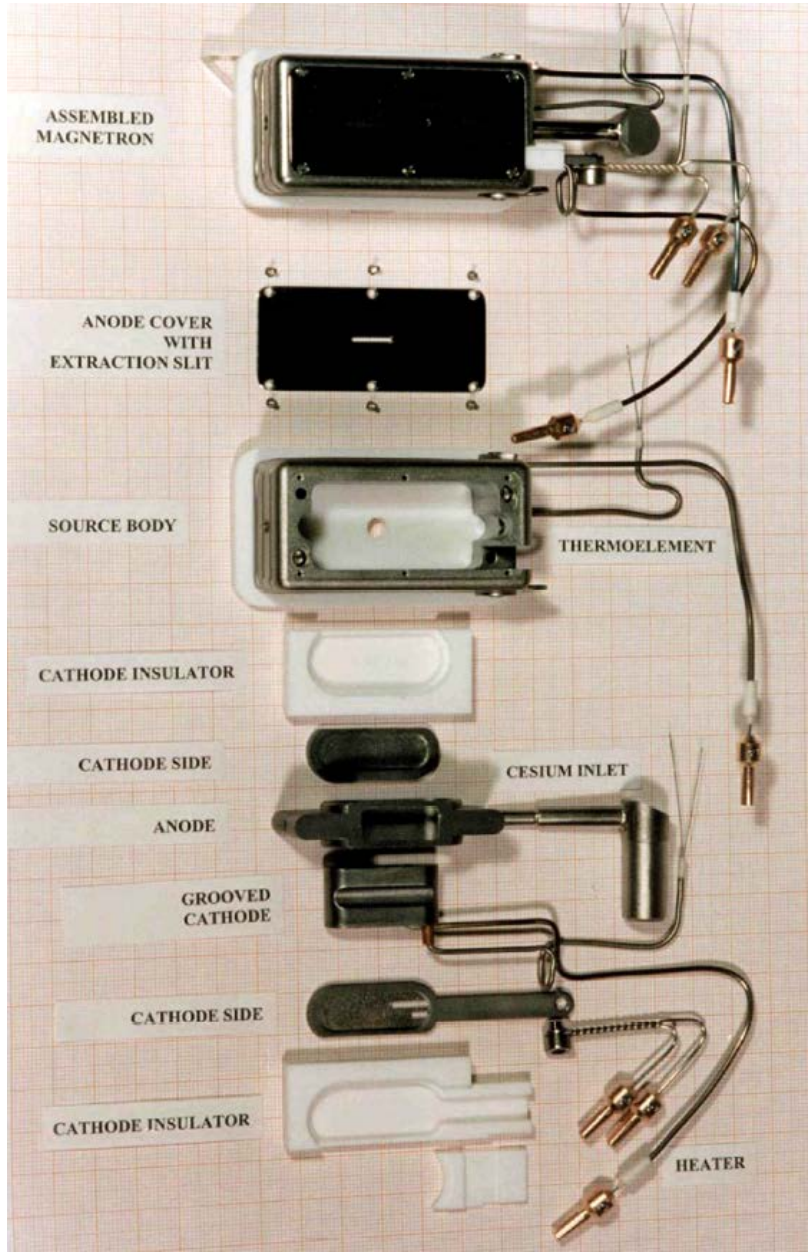

**Fig. 11:** The HERA magnetron components

### 5.8.4 Brookhaven National Laboratory (BNL), USA

Jim Alessi at Brookhaven further optimized the FNAL design [17]. He was able to reduce the discharge current, increase the extraction voltage and use permanent magnets. This highly reliable, very efficient magnetron, shown in Fig. 12, is still in use today. The operational parameters for this source are shown in Table 1.

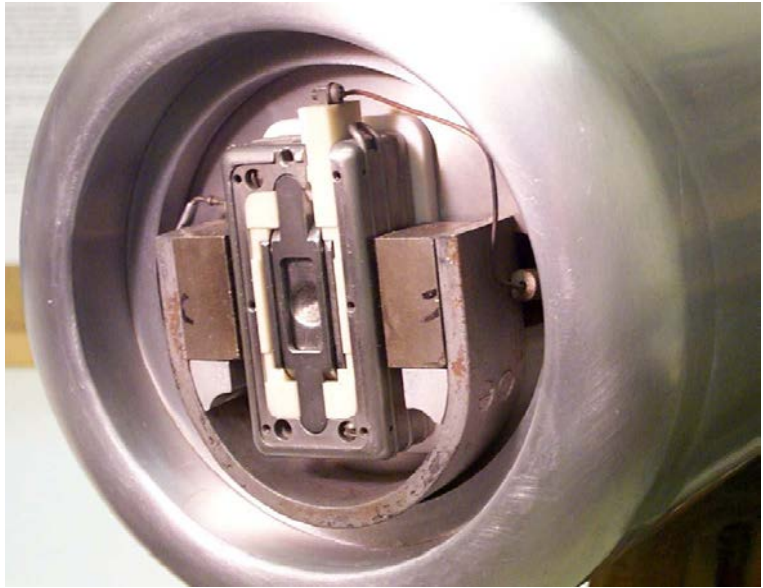

**Fig. 12:** The BNL magnetron with the anode cover plate and extract electrode removed

**Table 1:** The operational parameters of the BNL magnetron.

| | |
|---|---|
| H$^-$ beam current | 90–100 mA |
| Extraction voltage | 35 kV |
| Discharge voltage | 140–160 V |
| Discharge current | 8–18 A |
| Rep. rate | 7.5 Hz |
| Beam pulse width | 700 µs |
| Duty factor | 0.5% |
| Cs consumption | 0.5 mg h$^{-1}$ |
| H$_2$ gas flow | 3 mL min$^{-1}$ |
| R.m.s. emittance | 0.4 π mm mrad (normalized) |
| Lifetime | 270 days |

# 6 Penning surface plasma source

## 6.1 Introduction

This section gives more details specifically about Penning surface plasma negative ion sources. Examples from different operational particle accelerators around the world are given.

## 6.2 Construction

The Penning negative ion source (Fig. 13) has a small (approximately 10 mm × 5 mm × 5 mm) rectangular discharge region with a transverse magnetic field. The long sides of discharge are bounded by two cathodes, with the other four walls at anode potential, creating a 'quadrupole-like' electric field arrangement. The two cathodes are parallel jaws machined from a single piece of molybdenum to give good thermal conduction to the mounting flange. The magnetic field is oriented orthogonally to the cathode surfaces so that electrons emitted from the cathode are confined by the magnetic field lines and reflect back and forth between the parallel cathode surfaces. The primary anode is hollow and has holes through which hydrogen and caesium vapour are fed into the discharge. The extraction aperture plate is connected to the primary anode, which completes the window frame anode. The beam is extracted from the plasma through a slit in the extraction aperture plate by a high voltage applied to an extraction electrode. All electrodes are made of molybdenum.

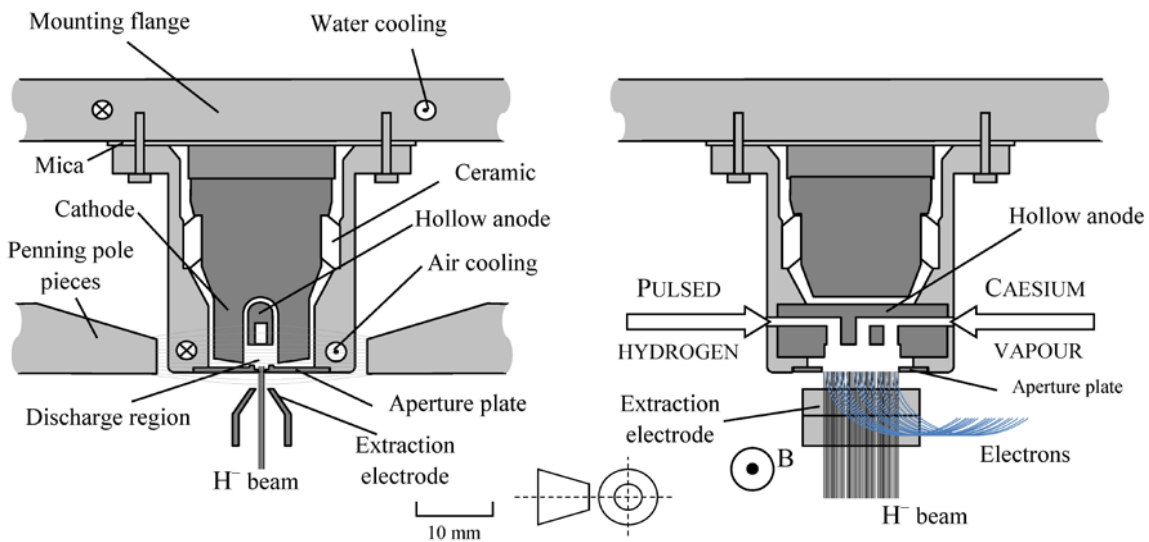

**Fig. 13:** Sectional schematic of a Penning H⁻ ion source

## 6.3 H⁻ production

Like with the magnetron, H⁻ ions are produced on the cathode surfaces and accelerated by the plasma sheath potential that exists next to the cathode. The plasma sheath potential is about 60 V. Unlike the magnetron, however, the cathode surface is not directly opposite the extraction aperture. The addition of ribs on the inside of the extraction aperture plate means that there is no direct line of sight between the cathode surface and the extraction aperture. Thus, it is impossible for the fast cathode-produced H⁻ to be extracted directly. Only H⁻ ions that have undergone resonant charge exchange with slow H⁻ ions (see section 3.4.2) are extracted, resulting in a beam with a lower energy spread than from the magnetron.

## 6.4 Temperature control

Like the magnetron discussed in the previous section, technically speaking the Penning negative ion source is a cold-cathode surface plasma Penning. Also, like the magnetron, 'cold-cathode' refers to the fact that the cathode is not a hot filament, but the cathode must still operate at elevated temperatures to maintain the Cs balance required for stable operation. The Cs-coated electrodes must also run at an elevated temperature to emit significant numbers of electrons. The electrodes generally run at a few hundred degrees Celsius, heated by the power of the discharge and temperature-controlled by some form of cooling system. From Eq. (3) the thermionic emission for Cs on Mo increases by 13 orders of magnitude between 300 K and 600 K. This is why Penning surface plasma sources will not run at room temperature.

## 6.5 Extraction

The Penning is the brightest $H^-$ ion source, with current densities at extraction above 1 A cm$^{-2}$ possible. The same points as mentioned in section 5.4 about the magnetron also apply to the Penning.

## 6.6 Caesium trapping

The same points as mentioned in section 5.6 about the magnetron also apply to the Penning.

## 6.7 Examples of negative Penning sources around the world

### 6.7.1 Budker Institute of Nuclear Physics (BINP), Novosibirsk, Russia

The $H^-$ Penning ion source was invented by Vadim Dudnikov, and developed with Gennadii Dimov and Yuri Belchenko at the Budker Institute of Nuclear Physics in the early 1970s at the same time as the magnetron source. Dudnikov reported pulsed $H^-$ currents up to 150 mA and that the source could run with duty cycles all of the way up to d.c. [14]. New versions of the Penning source continue to be developed at BINP by Belchenko and his team. In 2012 they reported $H^-$ beam currents of 25 mA d.c. [19].

### 6.7.2 Institute for Nuclear Research (INR), Moscow, Russia

An almost unmodified version of the original 1970s BINP source is still routinely used at INR. Two views of the INR Penning are shown in Fig. 14. The first shows the anode and cathode with the anode cover plate and extract electrode removed, and the second shows the ion source assembly.

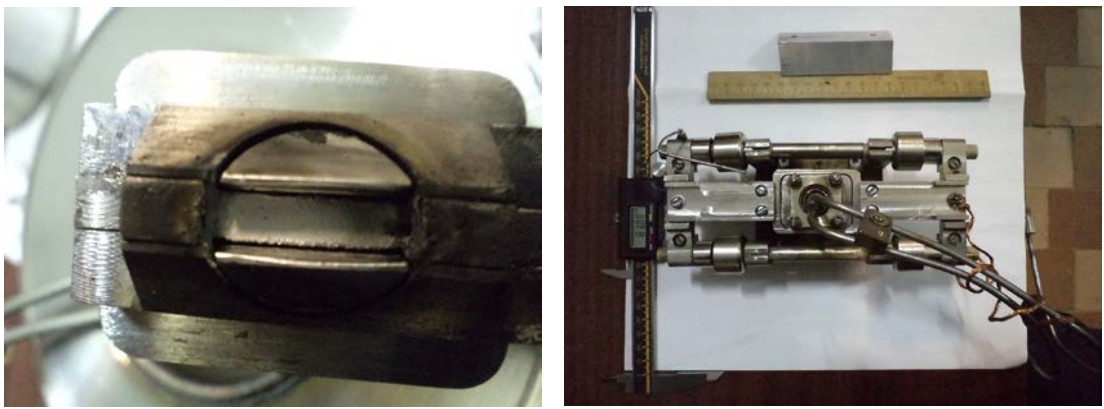

**Fig. 14:** The INR Penning $H^-$ ion source

### *6.7.3 Los Alamos National Laboratory (LANL), USA*

The Americans took the BINP Penning design and developed it further. In the 1990s Vernon Smith, Paul Allison and Joe Sherman at Los Alamos developed a scaled-up Penning source [20] shown in Fig. 15 that produced 120 mA H⁻ beam currents in 500 µs pulses at 60 Hz.

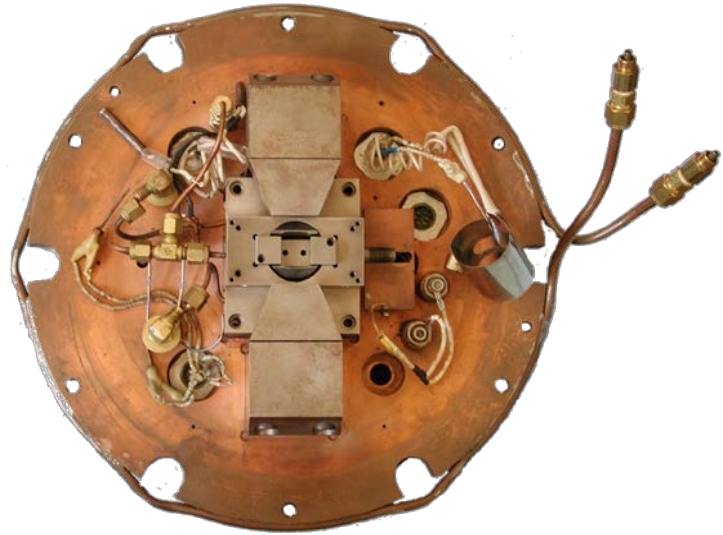

**Fig. 15:** The LANL scaled Penning H⁻ ion source with the anode cover plate and extract electrode removed

### *6.7.4 Rutherford Appleton Laboratory (RAL), UK*

In the early 1980s Peter Gear and Roger Bennett at RAL combined an early LANL Penning design with some of the aspects of the FNAL magnetron design. They used the FNAL 90° magnet, cold box and rear-loading design for use on the ISIS accelerator. The design has slowly evolved over the years. The present design used for ISIS operations is shown in Fig. 16.

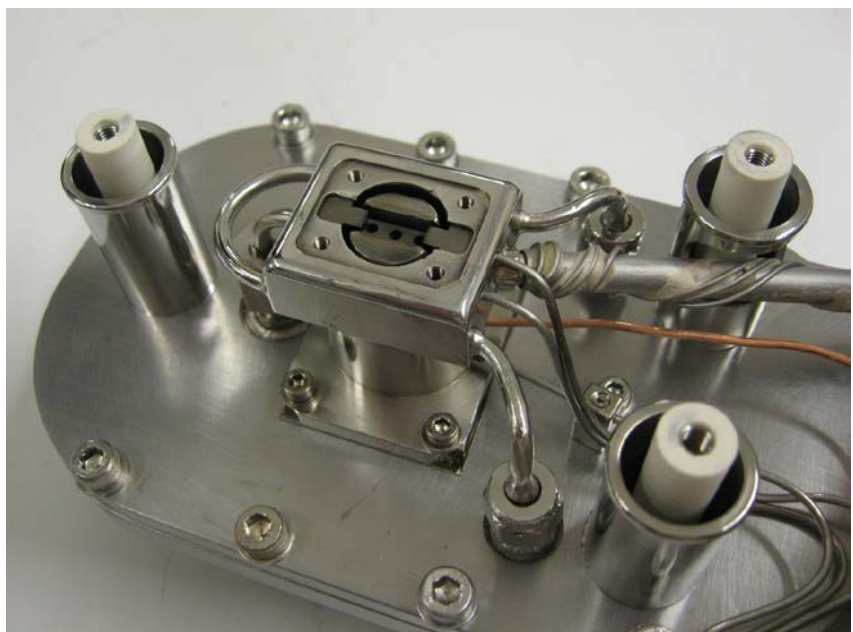

**Fig. 16**: The RAL Penning H⁻ ion source with the anode cover plate and extract electrode removed

The operational parameters for the ISIS source are shown in Table 2.

This source is currently under development at the Rutherford Appleton Laboratory by the author and his team, and 60 mA, 1 ms, 50 Hz pulses are now routinely produced [21].

**Table 2:** The operational parameters of the ISIS operational Penning source.

| H$^-$ beam current | 35 mA |
|---|---|
| Extraction voltage | 17 kV |
| Discharge voltage | 60–80 V |
| Discharge current | 55 A |
| Rep. rate | 50 Hz |
| Beam pulse length | 250 µs |
| Duty factor | 0.5% |
| Cs consumption | 3 mg h$^{-1}$ |
| H$_2$ gas flow | 20 mL min$^{-1}$ |
| R.m.s. emittance | 0.35 π mm mrad (normalized) |
| Lifetime | 25 days |

## 7 Discussion

### 7.1 Operating and tuning

All ion sources require a certain amount of tuning of the operating parameters for optimum performance. Surface plasma sources are no exception.

The main operating parameters are: hydrogen pressure, caesium vapour pressure, discharge current, magnetic field strength and the temperature of the electrodes. Secondary parameters such as timing and repetition rate have an effect on the transient hydrogen pressure and the degree of caesium coverage on the electrodes. Electrode cooling controls the temperature of the electrodes, which also controls the degree of caesium coverage.

Surface plasma ion sources are difficult to work with because their operation depends on optimum electrode surface temperature and optimum caesium coverage. Unfortunately, these optima are in the range of several hundred degrees Celsius and less than a monolayer, respectively, so it is impossible to just set all the parameters and turn the source on. The optima must be arrived at over time.

The optimum surface temperature must be arrived at by heating the electrodes (either with heaters or by the discharge itself). The optimum caesium coverage must be arrived at by allowing the flux of caesium onto the electrode surfaces to stabilize with the caesium sputtering/desorption rate from the surfaces.

When the optimum electrode temperature and caesium coverage have been reached, the source must be stabilized. This involves applying the correct amount of cooling to maintain the average electrode temperature, and the correct caesium oven temperature to maintain the flux onto the surface.

The caesiated Penning geometry operates in two distinct types of discharge mode: high-impedance (low-current), and low-impedance (high-current) modes. There is a sharp transition between the two modes that occurs when the electrodes are hot enough to maintain the correct caesium coverage. It is the low-impedance (high-current) discharge mode that is required for high-current $H^-$ ion production.

The two operating modes have very different operating conditions: the high-impedance mode has a discharge voltage of several hundred volts, whereas the low-impedance mode has a discharge voltage of about 60–70 V. When the source is cold and first switched on, it runs in high-impedance mode. It is impossible to produce a high-current plasma in high-impedance mode.

In normal running conditions, the source operates in low-impedance (high-current) mode. If the source drifts from its optimum conditions, it can slip back into high-impedance mode. The discharge current drops, the electrodes cool, the source becomes unstable and the discharge goes out. It can only be restarted by applying a high voltage and recovering the optimum surface temperatures and caesium coverage. This can make experimenting with different operating parameters quite time-consuming.

The best way to run surface plasma sources for operational accelerators (once they have started up) is to try and keep everything steady. Any changes to the operating parameters should be made one at a time and in very small steps with long pauses (hours or days) in between. This process is called tuning, and requires skill and patience.

## 7.2   Sputtering

In both magnetron and Penning ion sources, the discharge is in direct contact with the anode and cathode, so sputtering processes will eventually erode the electrode surfaces. This puts a fundamental limit on the lifetime of magnetron and Penning surface plasma sources. The degree of sputtering increases with the size of the projectile atom. Caesium atoms are therefore a major contributor to the total amount of sputtering that occurs.

Sputtering not only removes material from electrode surfaces but also redeposits it. This can cause build-up of sputtered material in the discharge volume. Figure 17 shows the inside of the plasma aperture plate electrode from the RAL Penning $H^-$ ion source after 26 days running. Significant erosion of the ribs either side of the aperture slit are visible. Deposited material is also visible all over the plate. The deposited material is molybdenum that has been sputtered from the cathode surfaces by bombardment with $Cs^+$ ions accelerated by the cathode sheath.

The sputtered material can flake off and bounce around inside the discharge volume, causing the discharge to appear unstable. The high-current discharge is a violent place and the flake debris can be broken up and vaporized; however, sometimes a flake can partially block the aperture slit, as shown in Fig. 17. This causes a step change in the beam current. If the blockage is small, it might be possible to increase the output current by tuning. It is also possible that a flake might short out the anode and cathode. If this happens, all beam is lost, and the source starts to cool. It is very rare, but possible, that a flake large enough to short out the extraction electrode is spat out of the aperture slit.

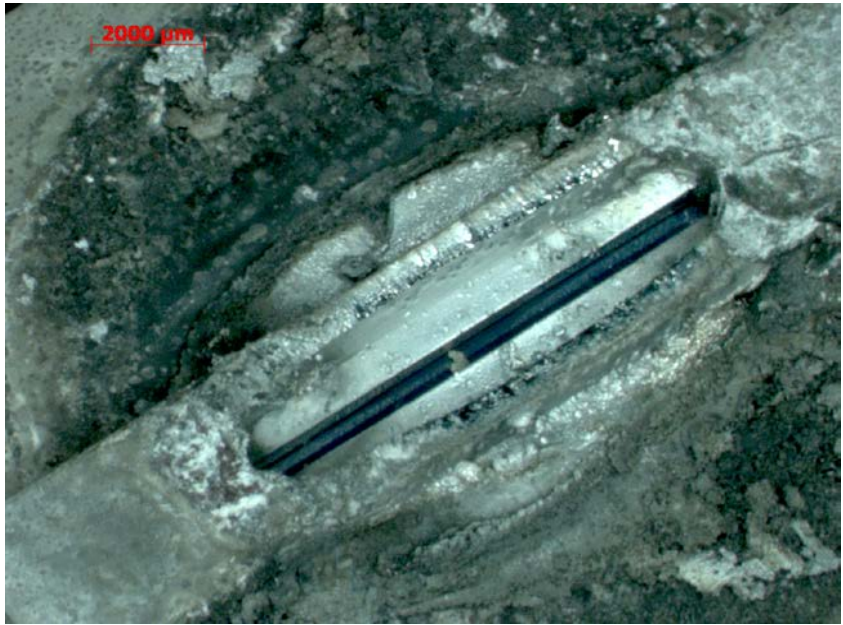

**Fig. 17**: The inside of the plasma aperture plate electrode from the RAL Penning H⁻ ion source after 26 days running.

### 7.3  Other failure modes

The other source failures are caused by failures of the ancillary equipment used to make the source work. These could be failures in power supplies, temperature controllers, heating, cooling, hydrogen delivery, caesium delivery, vacuum pumps, timing and control systems etc.

### 7.4  Lifetime comparison

Magnetrons are generally used for lower duty cycles and Penning sources for higher duty cycles. Table 3 shows the lifetimes of some of the different sources that have good operational data. On first inspection, the sources have very different lifetimes; however, when the plasma duty factors are taken into account, the lifetimes in integrated plasma days are similar.

**Table 3:** Comparison of the lifetimes of the different surface plasma sources.

|  | DESY magnetron | FNAL magnetron | BNL magnetron | RAL Penning |
|---|---|---|---|---|
| Discharge current (A) | 47 | 50 | 18 | 55 |
| Plasma pulse length (µs) | 75 | 80 | 700 | 800 |
| Rep. rate (Hz) | 6.25 | 15 | 7.5 | 50 |
| Plasma duty factor (%) | 0.047 | 0.12 | 0.525 | 4 |
| Lifetime (days) | 900 | 200 | 270 | 25 |
| **Lifetime (plasma days)** | **0.42** | **0.24** | **1.42** | **1.00** |

## Acknowledgements

Many thanks are due to Alan Letchford and David Findlay, who diligently proof-read this paper and gave excellent suggestions for improvement. Thanks also go to Scott Lawrie for reading before publication. Thanks also to Jim Alessi, Dan Bollinger and Jens Peters for pictures of their magnetron sources, and to Viktor Klenov for INR Penning source photographs.

## References

[1] A.W. Hull, *Phys. Rev.* **18** (1921) 31–57.
[2] J.E. Brittain, *Proc. IEEE* **98** (2010) 635–637.
[3] E.S. Lamar and O. Luhr, *Phys. Rev.* **44** (1933) L947–948.
[4] S.N. Van Voorhis *et al.*, *Phys. Rev.* **45** (1934) L492–493.
[5] L.R. Maxwell, *Rev. Sci. Instrum.* **2** (1931) 129.
[6] F.M. Penning, *Physica* **IV**(2) (1937) 71–75.
[7] B.L. Donnally, *Phys. Rev.* **159** (1967) 87.
[8] K.W. Ehlers *et al.*, *Nucl. Instrum. Methods* **22** (1963) 87–92.
[9] K.W. Ehlers, *Nucl. Instrum. Methods* **32** (1965) 309–316.
[10] G.P. Lawrence *et al.*, *Nucl. Instrum. Methods* **32** (1965) 357–359.
[11] L.E. Collins and R.H. Gobbett, *Nucl. Instrum. Methods* **35** (1965) 277–282.
[12] V.E. Krohn, Jr, *J. Appl. Phys.* **33** (1962) 3523–3525.
[13] Yu.I. Belchenko, G.I. Dimov and V.G. Dudnikov, *Nucl. Fusion* **14** (1974) 113.
[14] V.G. Dudnikov, Proc. 4th All-Union Conf. on Charged Particle Accelerators, Moscow, 1974 (Nauka, Moscow, 1975), vol. 1, pp. 323–325.
[15] K.N. Leung and K.W. Ehlers, *Rev. Sci. Instrum.* **53** (1982) 803.
[16] J.B. Taylor and I. Langmuir, *Phys. Rev.* **51** (1937) 753–760.
[17] J.G. Alessi, in *High Intensity and High Brightness Hadron Beams*, 20th ICFA Int. Beam Dynamics Workshop, AIP Conf. Proc. 642 (American Institute of Physics, Melville, NY, 2002), pp. 279–281.
[18] J. Peters, in *Negative Ions, Beams and Sources*, Proc. 1st Int. Symp., AIP Conf. Proc. 1097 (American Institute of Physics, Melville, NY, 2009), pp. 236–242.
[19] Yu.I. Belchenko *et al.*, in *Negative Ions, Beams and Sources*, Proc. 3rd Int. Symp., AIP Conf. Proc. 1515 (American Institute of Physics, Melville, NY, 2013), pp. 448–455.
[20] H.V. Smith *et al.*, *Rev. Sci. Instrum.* **65** (1994) 123.
[21] D.C. Faircloth *et al.*, in *Negative Ions, Beams and Sources*, Proc. 3rd Int. Symp., AIP Conf. Proc. 1515 (American Institute of Physics, Melville, NY, 2013), pp. 359–368.